\documentclass[prb,twocolumn,showpacs]{revtex4-1}
\usepackage{amsmath}
\usepackage{dcolumn}
\usepackage[english]{babel}
\usepackage{graphicx}
\usepackage[small,bf,hang,raggedright,FIGTOPCAP]{subfigure}
\usepackage{varioref}
\usepackage{epstopdf}
\usepackage{color}
\usepackage{eucal}
\input epsf
\usepackage{nicefrac}
\usepackage{IEEEtrantools}

\newcommand{\bq}{{\mathbf q}}
\newcommand{\bqp}{{\mathbf q'}}
\newcommand{\bqpp}{{\mathbf q''}}

\newcommand{\w}{\omega}

\begin{document}
\title{Ab initio variational approach for evaluating lattice thermal conductivity}
\author{Giorgia Fugallo, Michele Lazzeri, Lorenzo Paulatto and Francesco Mauri}
\affiliation{IMPMC, Universit\'e Pierre et Marie Curie, CNRS, 4 place Jussieu, F-75252 Paris, France\\}

\begin{abstract}
We present a first-principles theoretical approach for evaluating the lattice thermal conductivity  based on the exact solution of the Boltzmann transport equation. 
We use the variational principle and the conjugate gradient scheme, which provide us with an algorithm faster than the one previously used in literature and able to always converge
to the exact solution. Three-phonon normal and umklapp collision, isotope scattering and border effects are rigorously treated in the calculation. Good agreement with experimental data for diamond is found. Moreover we show that by growing more enriched diamond samples it is possible to achieve values of thermal conductivity up to three times larger than the commonly observed in isotopically enriched diamond samples with $99.93$\% C$^{12}$ and $0.07$ C$^{13}$.

\end{abstract}
\maketitle

\section{Introduction}

Thermal conductivity is one of the most important parameters used to characterize transport phenomena in solid state systems.
A predictive theory for evaluating thermal conductivity is essential for the design of new materials for efficient thermoelectric 
refrigeration and power generation\cite{matersoc} and it could help in understanding heat dissipation in micro- and nano-electronics devices.\cite{chen2}

When heat is mostly carried by lattice vibrations, such as in semiconductors and insulators, a correct theoretical prediction of thermal transport properties
cannot leave aside an accurate description of the phonon-phonon interactions and lifetimes.
These quantities are related to second and third order derivatives of the ground state energy with respect to atomic displacements.
Specifically the harmonic interatomic force constants determine phonon frequencies, group velocities and phonon populations while the anharmonic interatomic force constants
determine phonon scattering rates and linewidths.\\
A first microscopic description of the thermal conductivity in semiconductors and insulators has been formulated in 1929 by Peierls and it has
become known as Boltzmann Transport Equation (BTE). This equation involves the unknown perturbed population of a phonon mode and it
describes how the perturbation due to a gradient of temperature is balanced by the change in the phonon population due to scattering processes.
A good predictive theory requires then a good knowledge of i) the harmonic and anharmonic inter-atomic force constants (IFCs) and ii) the perturbed phonon population obtained as solution of the BTE.\\
Both these issues have non-trivial solutions.
The first issue can be addressed in the framework of Density Functional Perturbation Theory (DFPT)\cite{baroni:rev} evaluating the interatomic force constants fully ab initio using the ``2n+1'' theorem\cite{debe:prl,michele,tedeschi}.
An efficient implementation of this method, extended to metallic systems, exists in the Quantum EPRESSO package\cite{qe} for zone-centered modes \cite{michele}.
A generalization for metallic systems and arbitrary phonons has recently been developed and implemented in the the Quantum-ESPRESSO package \cite{paulatto}. 
 The second issue, lying in solving the BTE equation exactly,  is due to the complexity of the scattering term. The change in the phonon population numbers of each single state involved in the 
scattering term depends, in turn, on the change in the occupation number of the other states involved. \\
Several theoretical studies, instead of attempting to solve the BTE, employ  a common approximation, namely the single mode phonon relaxation time approximation (SMA) \cite{callaway:pr,klemens,ziman}. 
This approximation describes rigorously the depopulation of the phonon states but not the corresponding repopulation, which is assumed to have no memory of the initial phonon distribution. 
 The momentum conserving character of the normal (N) processes gives then rise to a conceptual inadequacy of the SMA description and its use becomes questionable in particular in the range of low temperatures where the umklapp (U) processes are frozen out and N
processes dominate the phonon relaxation \cite{guyer:pr1}. \\
Improved approximate techniques involve the use of a variational procedure \cite{petterson:prb43, petterson:jpc}.
In such a kind of approach, originally introduced by Hamilton and Parrott \cite{parrot:prb178},
the thermal conductivity is found by variationally optimizing a trial function 
describing the non-equilibrium phonon distribution function.
Unfortunately the less the system is symmetric and isotropic the more the result and the accuracy will be affected by the form adopted for the trial function. \\

A first approach to solve exactly the linearized BTE has been introduced in the 90s by Omini and Sparavigna \cite{sparavigna:physb}.
The numerical solution evaluated on a reciprocal space discrete grid is obtained via a self-consistent iterative procedure, but
as indicated by the authors \cite{sparavigna:physb} there is no general proof that convergence will always be obtained with this approach. In particular the method shows an instability that 
prevents it from reaching the exact solution in the range where N phonon scattering processes dominate and the other scattering processes are weak.
Nevertheless until now the Omini Sparavigna (OS) iterative procedure has represented the only numerically exact method used to solve the BTE and evaluate the thermal conductivity with \cite{broido:apl, broido:prb80,broido:prb84,marzari:prl}  and without \cite{sparavigna:nc,sparavigna:prb53} IFCs from ab initio approaches. The method scales as the square of the number of grid points and it requires very dense grids to converge the thermal conductivity. As a consequence, the time required to solve the BTE could dominate over the time required to compute the IFCs even when these are evaluated by first principles.


In this paper we present a new approach for solving exactly the linearized BTE. This method joins together the variational principle and the resolution on a discrete grid. More specifically by using the variational principle and the conjugate gradient method, we present a stable algorithm, faster than the one previously proposed and able to always  converge to the exact solution. \\
In particular the mathematical stability assures the possibility to use the present method for evaluating the thermal conductivity in all the possible ranges of temperatures, without  the problems \cite{sparavigna:physb} found by the previous method. These properties assure the flexibility of the present approach in treating any structure without any a-priori knowledge. \\
Moreover, even in the case where both of the methods are stable, the present scheme assures to reach the convergence one order of magnitude more rapidly than the OS, opening the possibility to treat more complex systems.

As a first application we use this algorithm for studying the lattice thermal conductivity in naturally occurring and isotopically enriched diamond. Diamond thermal conductivity is the highest known among bulk materials. At room temperature its value is more than an order of magnitude higher than in other semiconductor materials, exceeding 3000 W/m-K\cite{onn,olson:prb,wei}. Diamond, and in general carbon systems, have strong covalent bonding and light atomic masses, which lead to high phonon frequencies, high acoustic velocities, and a very small phase space for Umklapp scattering when compared with other common semiconductors. As a consequence, large amounts of heat are transferred by acoustic phonons with high velocities, giving these systems their high values of thermal conductivity\cite{berman:prb45,broido:prb80,broido:prb87,slack}. Weak Umklapp phonon scattering makes the system very sensitive to small changes in the isotopic content at low temperatures. Different data are available for a large temperature range and for a wide range of C$^{13}$ isotope concentrations\cite{onn,berman:prb45,olson:prb,wei,vandersande,anthony1,anthony2,anthony3}. In our case this has the double advantage of enabling us to : i) test the stability of the present algorithm with respect to the OS method, even in cases where N scattering processes are dominant with respect to the other scattering events such as in isotopically enriched diamond; and, physically more interesting: ii)  give a theoretical limit based on the exact solution of the BTE of the maximum lattice thermal conductivity reachable in isotopically pure diamond samples. 

\section{Boltzmann trasport equation}
\begin{figure}[htp]
\includegraphics[width= 0.48\textwidth, angle=0]{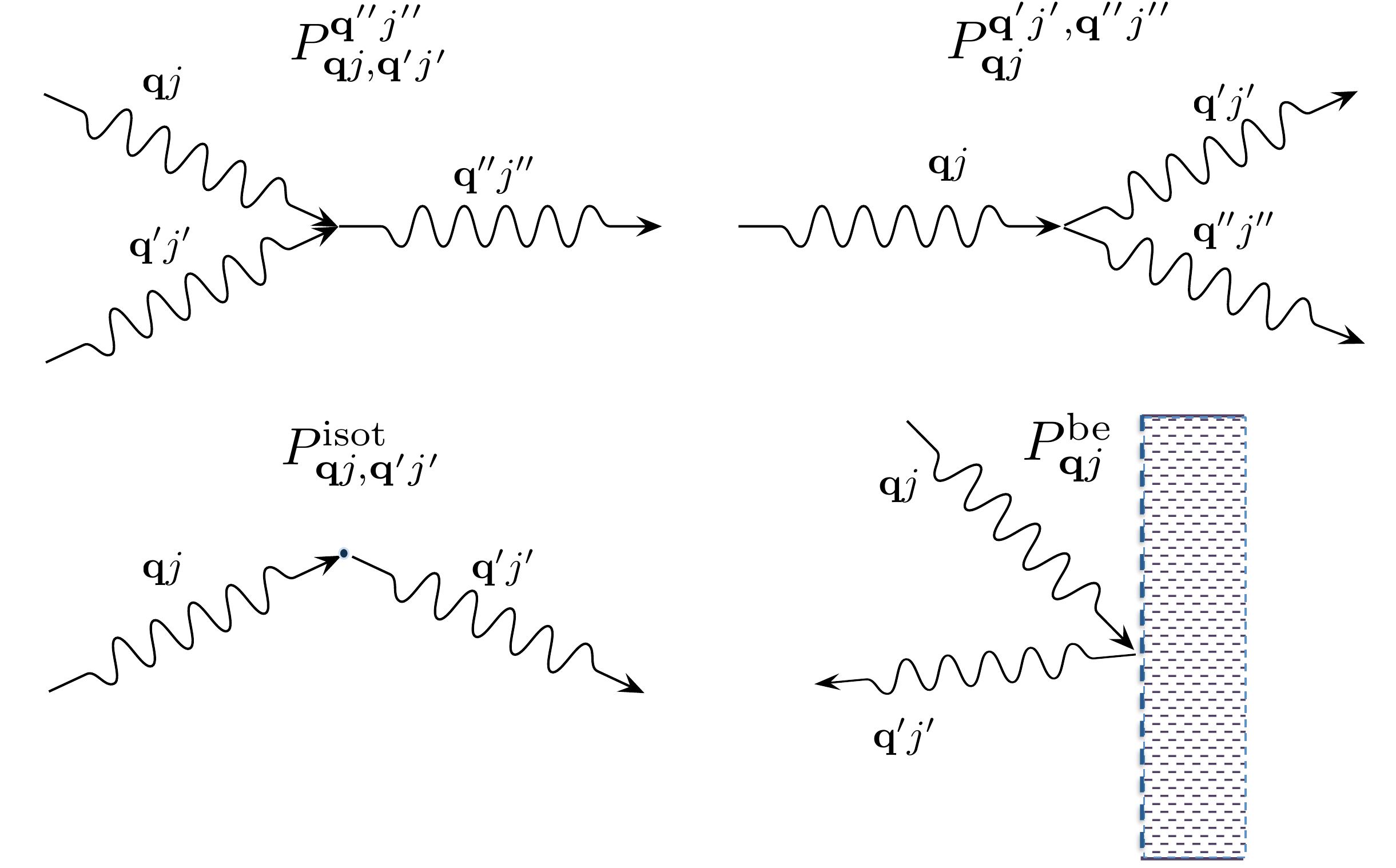}
\caption[Anharmonic Feynman diagrams]{Phonon scattering processes in a finite size anharmonic crystal in presence of isotopic impurities.}\label{classEv}
\end{figure}

When a gradient of temperature $\nabla T$ is established in a system, a subsequent 
heat flux will start propagating in the medium. Without loss of generality we assume the gradient of temperature to be along 
the direction $x$.
 The flux of heat, collinear to the temperature gradient, can be written in terms of phonon energies $\hbar\omega_{\mathbf{q}j}$, phonon
group velocities $c_{\bq j} $ in the $x$ direction, and the perturbed phonon population $n_{\bq j}$:
\begin{equation}
\frac{1}{N_0 \Omega}\sum_{\bq j}\hbar \w_{\bq j}c_{\bq j}n_{\bq j}=-k  \frac{\partial T}{ \partial x}
\end{equation}
On the l.h.s $\omega_{\mathbf{q}j }$ is the
angular frequency of the phonon mode with vector $\mathbf{q}$ and branch index $j$, $\Omega$ is the volume of 
the unit cell and the sum runs over a uniform mesh of $N_0$ $\mathbf{q}$ points. 
On the r.h.s. $k$ is the diagonal component of the thermal conductivity in the temperature-gradient direction. 
 Knowledge of the perturbed phonon population allows heat flux and subsequently thermal conductivity to be evaluated. \\
Unlike phonon scattering by defects, impurities and boundaries, anharmonic scattering represents an intrinsic resistive 
process and in high quality samples, at room temperature, it dominates the behaviour of lattice thermal conductivity balancing the perturbation due to the gradient of temperature.
The balance equation, namely the Boltzmann Transport Equation (BTE), formulated in 1929 by Peierls \cite{peierls1} is:
\begin{equation}
-c_{\bq j}\frac {\partial T} {\partial x} \left(\frac{\partial n_{\bq j}}{\partial T}\right)
    +\left.\frac{\partial n_{\bq j}}{\partial t}\right|_{scatt}=0
\label{BTE1}
\end{equation}
with the first term indicating the phonon diffusion due to the temperature gradient and the second term the scattering rate due to all the scattering processes.
This equation has to be solved self consistently. In the general approach \cite{ziman}, for small perturbation from the equilibrium,  the temperature gradient of the perturbed phonon population is replaced with the temperature gradient of the equilibrium phonon population $\partial n_{\bq j} / \partial T = \partial \bar{n}_{\bq j} / \partial T $ where $\bar{n}_{\bq j} = (e^{\hbar \w_{\bq j} /k_BT} - 1)^{-1}$; while for the scattering term it can be expanded about its equilibrium value in terms of a first order perturbation $f^{\mathrm{EX}}$:
 \begin{equation}
 n_{\bq j} \simeq \bar{n}_{\bq j}+\bar{n}_{\bq j}(\bar{n}_{\bq j}+1) \frac{\partial T}{\partial x}f^{\mathrm{EX}}_{\bq j}
\label{LINEAR-N}
 \end{equation}
The linearized BTE 
 can then be written in the following form \cite{sparavigna:prb66}:
\begin{eqnarray}
-c_{\bq j}\left(\frac{\partial \bar{n}_{\bq j}}{\partial T}\right) &=&  
  \sum_{\bqp j',\bqpp j''}\Big[ P_{\bq j,\bqp j'}^{\bqpp j''}(f^{\mathrm{EX}}_{\bq j}+f^{\mathrm{EX}}_{\bqp j'}-f^{\mathrm{EX}}_{\bqpp j''}) \nonumber  \\
  &+&\frac{1}{2} P^{\bqp j',\bqpp j''}_{\bq j} (f^{\mathrm{EX}}_{\bq j}-f^{\mathrm{EX}}_{\bqp j'}-f^{\mathrm{EX}}_{\bqpp j''} )\Big] \nonumber \\
   &+& \sum_{\bqp j'}  P^{\mathrm{isot}}_{\bq j,\bqp j'}  (f^{\mathrm{EX}}_{\bq j} - f^{\mathrm{EX}}_{\bqp j'}) \nonumber \\
   &+& P^{\mathrm{be}}_{\bq j} f^{\mathrm{EX}}_{\bq j}
\label{BTE2}
\end{eqnarray}
where the sum  on $\bqp$ and $\bqpp$ is performed in the Brillouin Zone (BZ). The $\mathrm{EX}$ superscript of the first order perturbation $f^{\mathrm{EX}}$
denotes the exact solution of the BTE, to be distinguished from the approximated solutions that  we will discuss later.\\
In Eq. \ref {BTE2}
the anharmonic scattering processes as well as the scattering with the isotopic impurities and the border
effect are considered. 
More specifically (see Fig.\ref{classEv}) $P_{\bq j,\bqp j'}^{\bqpp j''}$  is the scattering rate at the equilibrium  of  a  process  
where a phonon mode $\bq j$ scatters by absorbing another mode $\bqp j'$ to generate a third phonon mode $\bqpp j''$. While $P^{\bqp j',\bqpp j''}_{\bq j}$ is the scattering rate at the equilibrium of a process where a phonon mode $\mathbf{q}j$ decays in two modes $\mathbf{q}'j'$ and $\mathbf{q}''j'' $. 

The two scattering rates have the forms:
\begin{eqnarray}
P^{\bqpp j''}_{\bq j,\bqp j'}&{=}& \frac{2 \pi}{N_0 \hbar^2} \sum_{\mathbf{G}}
	  |V^{(3)}(\bq j,\bqp j',{-}\bqpp j'')|^2  \nonumber \\
	&&    \bar{n}_{\bq j}\bar{n}_{\bqp j'}(\bar{n}_{\bqpp j''}+1) \delta_{\bq{+}\bqp {-}\bqpp, \mathbf{G}}\nonumber \\
	&&  \delta(\hbar \w_{\bq j} +\hbar \w_{\bqp j'}-\hbar \w_{\bqpp j''}) \label{coal}  
\end{eqnarray}
\begin{eqnarray}
P^{\bqp j',\bqpp j''}_{\bq j}&{=}& \frac{2 \pi}{N_0 \hbar^2 } \sum_{\mathbf{G}}
	    |V^{(3)}(\bq j,{-}\bqp j',{-}\bqpp j'')|^2 \nonumber \\
        &&       \bar{n}_{\bq j}(\bar{n}_{\bqp j'}{+}1)(\bar{n}_{\bqpp j''}{+}1)\delta_{\bq{-}\bqp {-}\bqpp, \mathbf{G}} \nonumber \\ 
        &&   \delta(\hbar \w_{\bq j}-\hbar \w_{\bqp j'}-\hbar \w_{\bqpp j''} )\label{dec} 
\end{eqnarray}
 with $ {\mathbf G}$ the reciprocal lattice vectors.
In order to evaluate  them 
it is necessary to compute the third derivative $V^{(3)}$
of  the total energy of the crystal $\mathcal{E}^{tot}(\{u_{s \alpha} (\mathbf{R}_l) \})$,
with respect to the atomic displacement $u_{s \alpha} (\mathbf{R}_l)$, 
from the equilibrium position, of the $s$-th atom, 
along the $\alpha$ Cartesian coordinate 
in the crystal cell identified by the lattice vector $\mathbf{R}_l$ :
\begin{equation}
V^{(3)}(\mathbf{q} j,\mathbf{q}' j',\mathbf{q}'' j'')= \frac{\partial^3 \mathcal{E}^{cell}}
                                                    {\partial X_{\mathbf{q} j},\partial X_{\mathbf{q}' j'},\partial X_{\mathbf{q}'' j''}}
\end{equation}
 where $\mathcal{E}^{cell}$ is the energy per unit cell. The non-dimensional quantity $X_{\mathbf{q} j}$ is defined by 
\begin{equation}
X_{\mathbf{q} j}= \frac{1}{N_0}\sum_{l,s,\alpha} \sqrt{\frac{2 M_s \omega_{\bq j}} {\hbar}} z^{s \alpha^*}_{\bq j}  u_{s \alpha }(\mathbf{R}_l) 
e^{-i\mathbf{q}\cdot \mathbf{R}_l}
\end{equation}
$z^{s \alpha}_{\mathbf{q}j} $ being the orthogonal phonon eigenmodes normalized on the unit cell and $M_s$ the atomic masses. \\
The rate of the elastic scattering with isotopic impurities (see Fig.\ref{classEv}) has the form \footnote{we symmetrised the form reported in Ref. \onlinecite{sparavigna:nc}}:
\begin{eqnarray}
  P_{\bq j,\bqp j'}^{\mathrm{isot}} & = & \frac{\pi}{2 N_0} \omega_{\bq j}\omega_{\bqp j'}  
                   \left[ \bar{n}_{\bq j} \bar{n}_{\bqp j'} + \frac{\bar{n}_{\bq j} + \bar{n}_{\bqp j'}} {2} \right ] \nonumber \\
                 & &\sum_{s} g^{s}_{2}   \left|  \sum_{\alpha} z^{s \alpha^*}_{\mathbf{q}j} \cdot z^{s \alpha}_{\bqp j'}  \right|^2 \delta (\omega_{\bq j}- \omega_{\bqp j'}) \label{isot1}
\end{eqnarray}
where $g^s_2 = \frac{(M_s - \langle  M_s\rangle)^2}{ \langle M_s \rangle^2 }$ is the average over the mass distribution of the atom of type $s$.
In presence of two isotopes $M_s$ and $M_{s'}$ it can be written in terms of the concentration $\epsilon$ and mass change $\Delta M_s= M_{s'} - M_s$ :
\begin{equation}
 g^s_2=  \epsilon(1-\epsilon)  \frac{\left | \Delta M_s\right |}{ \langle M_s \rangle} 
\label{g2expr}
\end{equation}
with  $\langle M_s \rangle = M_s + \epsilon \Delta M_s$.\\
Eventually, in a system of finite size, $P_{\bq j}^{\mathrm{be}} $ describes the reflection of a phonon from the border (see Fig.\ref{classEv}):
\begin{equation}
P_{\bq j}^{\mathrm{be}} = \frac{c_{\bq j}}{LF}\bar{n}_{\bq j}(\bar{n}_{\bq j}+1) 
\label{bord}
\end{equation}
where $L$ is the Casimir length of the sample and $F$ a correction factor depending 
on the width to length ratio of the sample. Following the literature \cite{sparavigna:prb65,broido:underp,born:book} the border scattering is treated in the relaxation time approximation
and it results in a process in which a phonon from a specific state($\mathbf{q} j$) 
is reemitted from the surface contributing only to the equilibrium distribution. \\
For the sake of clarity we will contract from here on the vector $\mathbf{q}$ and branch index $j$ in a single mode index $\nu$. 
The BTE of Eq. \ref{BTE2} can be written as  a linear system in matrix form:
\begin{equation}
\mathbf{A} \mathbf{f}^{\mathrm{EX}}=\mathbf{b}
\label{linearsyst}
\end{equation}
with the vector $b_{\nu'} =-c_{\nu'}\hbar \w_{\nu'} \bar{n}_{\nu'}(\bar{n}_{\nu'}+1)  $ and the matrix
\begin{multline}
\hspace{-0.42cm}
A_{\nu,\nu'}{=} \left[{\sum_{\nu'',\nu'''}}  \left(P^{\nu''}_{\nu,\nu'''} {+}\frac{ P_{\nu''',\nu''}^{\nu}}{2} \right) {+}\sum_{\nu''} P^{\mathrm{isot}}_{\nu,\nu''} {+} P^{\mathrm{be}}_{\nu} \right ] \delta_{\nu,\nu'}+\\
                - {\sum_{\nu''}} \left(  P^{\nu'}_{\nu,\nu''} -P^{\nu''}_{\nu,\nu'}+ P_{\nu',\nu''}^{\nu} \right )    + P^{\mathrm{isot}}_{\nu,\nu'} 
\label{ourA}
\end{multline}
where we have used $P^{\nu', \nu''}_{\nu}=P_{\nu', \nu''}^{\nu}$ from the detailed balance condition $   \bar{n}_{\nu}(\bar{n}_{\nu'}+1)(\bar{n}_{\nu''}+1) = (\bar{n}_{\nu}+1)\bar{n}_{\nu'}\bar{n}_{\nu''}$ (valid under the assumption  $\hbar \w = \hbar \w' + \hbar \w''$).
In this form the matrix is symmetric and positive semi-definite  (see Appendix A for demonstrations) and it can be decomposed in $\mathbf{A} = \mathbf{A}^{\mathrm{out}} +\mathbf{A}^{\mathrm{in}} $,
where
\begin{eqnarray}
A^{\mathrm{out}}_{\nu,\nu'} &=& \frac{\bar{n}_{\nu}(\bar{n}_{\nu} +1)} {\tau^{\mathrm{T}}_{\nu}}\delta_{\nu,\nu'} \\
A^{\mathrm{in}}_{\nu,\nu'} &=&  -  \sum_{\nu''} \left(  P^{\nu'}_{\nu,\nu''} -P^{\nu''}_{\nu,\nu'}+ P_{\nu',\nu''}^{\nu} \right )    + P^{\mathrm{isot}}_{\nu,\nu'} 
\end{eqnarray}
$\tau^{\mathrm{T}}_{\nu}$ being the phonon relaxation time (see Appendix B).  The $\mathbf{A}^{\mathrm{out}}$ diagonal matrix describes the depopulation of phonon states due to the scattering mechanisms while the
$\mathbf{A}^{\mathrm{in}}$ matrix describes their repopulation due to the incoming scattered phonons.
\\
The solution of the linear system in Eq. \ref{linearsyst} is obtained formally by inverting the matrix ${\mathbf A}$.
\begin{equation}
{\mathbf f}^{\mathrm{EX}} =   \frac{1}{\mathbf{A}}  {\mathbf b}
\label{generalSol}
\end{equation}
and subsequently the thermal conductivity will be evaluated as:
\begin{eqnarray}
k &= & \lambda {\mathbf b} \cdot {\mathbf f}^{\mathrm{EX}} \\ \nonumber
&=& - \frac{\hbar}{N_0\Omega  k_B T^2}\sum_{\nu}c_{\nu}
				    \w_{\nu} \bar{n}_{\nu}(\bar{n}_{\nu}+1) f_{\nu}^{\mathrm{EX}}
\label{definizione1}
\end{eqnarray}
with $\lambda= 1 /(N_0\Omega k_B T^2)$.

\section{Solutions of the Boltzmann transport equation}
The complexity of the BTE lies in the need of explicitly computing, storing and inverting the large matrix $\mathbf{A}$.
In the SMA the BTE is solved for the $n_{\nu}$ neglecting the role of the repopulation by means setting $\mathbf{A}^{\mathrm{in}}$ to zero
\begin{equation}
{\mathbf f}^{\mathrm{SMA}} =\frac{1}{ \mathbf{A}^{\mathrm{out}}}  {\mathbf b}
\end{equation}
Storing and inverting $\mathbf{A}^{\mathrm{out}}$ is trivial due to its diagonal form.
The lattice thermal conductivity in SMA is then 
\begin{equation}
k^{\mathrm{SMA}}=\lambda \mathbf{b} \cdot \mathbf{f}^{\mathrm{SMA}}=\frac{\hbar^2}{N_0\Omega k_B T^2}\sum_{\nu}c^2_{\nu}
				    \w^2_{\nu} \bar{n}_{\nu}(\bar{n}_{\nu}+1)\tau^{\mathrm{T}}_{\nu}.
\label{k-sma}
\end{equation}
Such approximation is exact if the repopulation loses memory of the initial phonon distribution and if it is proportional to the equilibrium population of $\nu$.
It remains anyway a good approximation if the repopulation is isotropic.
An exact solution of Eq. \ref{linearsyst}, that does not imply either storing or the explicit inversion of matrix $\mathbf{A}$, 
 has been proposed by Omini
and Sparavigna  \cite{sparavigna:nc} by converging with respect to the iteration $i$ the following:
\begin{equation}
\mathbf{f}_{ i+1} =\frac{1} {\mathbf{A}^{\mathrm{out} } } \mathbf{b} - \frac{1} {\mathbf{A}^{\mathrm{out} } } \mathbf{A}^{\mathrm{in}}  \mathbf{f}_{i}
\label{os1}
\end{equation}
with the iteration zero  consisting in the SMA $\mathbf{f}_0=\mathbf{f}^{\mathrm{SMA}}$.
This procedure requires, as for the SMA, only the trivial inversion of the diagonal matrix $\mathbf{A}^{\mathrm{out}}$. Instead of storing and inverting $\mathbf{A}$, it just requires the evaluation of $\mathbf{A}^{\mathrm{in}}\:\mathbf{f}_{i}$, at each iteration $i$ of the OS method, which is an operation computationally much less demanding, .\\
Once the convergence is obtained the thermal conductivity is evaluated by:
\begin{equation}
k^{\mathrm{NV}}(\mathbf{f}_i)=\lambda \mathbf{b}\cdot \mathbf{f}_{i}
\label{kOS}
\end{equation}
From a mathematical point of views the OS iterative procedure 
can be written as a geometric series:
 \begin{equation}
\mathbf{f}_{ i} = \sum_{j=0,i} \left(-\frac{1}{\mathbf{A}^{\mathrm{out}}}  \mathbf{A}^{\mathrm{in}}\right)^{j} \frac{1}{\mathbf{A}^{\mathrm{out}}} \:  \mathbf{b}
\label{os2}
\end{equation}
thus only if  the absolute value of the ratio  $\left((\mathbf{A}^{\mathrm{out})^ {-1}}  \mathbf{A}^{\mathrm{in}}\right) $is smaller than one the series converges to a solution of 
the linear system in Eq. \ref{linearsyst} . \\
An alternative approach consists in using the properties of the matrix ${\mathbf A} $ (see Appendix A) to find the exact solution of the linearized BTE, via the variational principle.
Indeed the solution  of the BTE is the vector $\mathbf{f}^{\mathrm{EX}}$ which makes  stationary the quadratic form \cite{klemens,parrot:prb178}
\begin{equation}
\mathcal{F}(\mathbf{f}) =\frac{1}{2} {\mathbf f} \cdot{\mathbf A} {\mathbf f}- {\mathbf b} \cdot {\mathbf f}
\label{fform}
\end{equation}
for a generic vector $\mathbf{f}$. Since $\mathbf{A}$ is positive the stationary point is the global and single minimum of this functional.
One can then define a variational conductivity functional: 
\begin{equation} 
k^\mathrm{V}(\mathbf{f}) = - 2 \lambda \mathcal{F}({\mathbf f})
\label{quadratic}
\end{equation}
that has the property $k^\mathrm{V}(\mathbf{f}^{\mathrm{EX}})=k$ while any other value of $k^{\mathrm{V}}(\mathbf{f})$  underestimates $k$.
In other words, finding the minimum of the quadratic form is equivalent to maximizing the thermal conductivity functional. 
As a consequence an error $\delta \mathbf{f}= \mathbf{f} - \mathbf{f}^{\mathrm{EX}}$  results in an error in conductivity, linear in $\delta \mathbf{f}$ if  the functional is written in Eq. \ref{kOS} form, and quadratic  if  the variational form (Eq.   \ref{quadratic}) is used.

In literature  \cite{parrot:prb178}, due to the complexity of the numerical calculations, the variational scheme was used together with a trial function for describing  the non-equilibrium phonon distribution function affecting then the accuracy of the final result with the form of the specific probe function chosen.
In our scheme we avoid the use of trial function and we solve Eq. \ref{linearsyst} on a grid (as in OS procedure) by using the conjugate gradient method \cite{num-rec}, as reported in Appendix C, to obtain the exact solution of the BTE equation.
In order to speed up the convergence of the conjugate gradient we take advantage of the diagonal and dominant role of $\mathbf{A}^{\mathrm{out}}$ and we use a preconditioned conjugate gradient. Formally, this corresponds to use in the minimization the rescaled variable:
\begin{equation}
\tilde{{\mathbf f}} = \sqrt{{\mathbf A^{\mathrm{out}}}} {\mathbf f}
\label{prec1}
\end{equation}
and then, with respect to this new variable, minimize the quadratic form $\tilde{\mathcal{F}}(\tilde{\mathbf{f}}) = \mathcal{F}(\mathbf{f})$ where:
\begin{equation}
\tilde{\mathcal{F}}( \tilde{\mathbf{f}}) =\frac{1}{2} \tilde{\mathbf{f}}\cdot \tilde{\mathbf{A}} \tilde{\mathbf{f}}- \tilde{\mathbf{ b}}\cdot\tilde{\mathbf {f}}
\label{tildefform}
\end{equation}
and  
\begin{eqnarray}
\tilde{{\mathbf A}} &=&\frac{1}{ \sqrt{{\mathbf A^{\mathrm{out}}}}} {\mathbf A}\frac{1}{ \sqrt{{\mathbf A^{\mathrm{out}}}}} \label{prec2} \\
\tilde{{\mathbf b}} &=&\frac{1}{ \sqrt{{\mathbf A^{\mathrm{out}}}}} {\mathbf b} \label{prec3}
\end{eqnarray}
Notice that $\tilde{\mathbf{f}}^{\mathrm{SMA}}=\tilde{\mathbf{b}}$. The square root evaluation of $\mathbf{A}^{\mathrm{out}}$ is trivial due to its diagonal form.
The computational cost per iteration of the conjugate gradient scheme is equivalent to the OS one, but it always converges and requires a smaller number of iterations.

\section{Computational Details}
In order to compute the thermal conductivity the only input required are the second and third order IFCs.
Both of them were calculated by using the Quantum ESPRESSO package\cite{qe} within a linear response approach\cite{baroni:rev,debe:prl,michele,tedeschi} following the method explained by Paulatto et al. \cite{paulatto}.
The first BZ is discretized into a uniform grid of ${\mathbf q}$ points centered at $\Gamma$, in such a way that if $\bq$ and $\bqp$ belong to the mesh also $\bq \pm \bqp$ belongs to the mesh, assuring a perfect momentum conservation.
 At any ${\mathbf q}$ the phonon frequencies are 
evaluated from the second order force constants and the phonon group velocities are computed from the derivative of the phonon dispersion $\partial \omega / \partial {\mathbf q}$, using the Hellmann-Feynman theorem and obtaining the following velocity matrix directly from the Dynamical matrix $\mathbf{\mathcal{D}}$:
\begin{equation}
C_{j j'} = \sum_{\alpha \alpha' s s'} \frac{1}{2 \sqrt{M_s M_{s'}} \omega_{\bq j} }  z^{s \alpha^*}_{\bq j}  \frac{\partial \mathcal{D}^{\alpha \alpha'}_{s s'}}{ \partial q_x}   z^{s' \alpha'}_{\bq j'}
\end{equation}
In the non-degenerate case $c_{\bq j}=C_{j j} $ while in the degenerate one we use the phonon polarization vectors that diagonalize the matrix in the degenerate subspace.
To compute the scattering rates, the BZ is again discretized into a grid of ${\mathbf q'}$ points centered in $\Gamma$.
The delta function for the energy conservation is replaced by a Gaussian
\begin{equation}
 \delta(\hbar \omega)=\frac{1} {\sqrt{\pi}  \sigma} \exp{(-(\hbar \omega/ \sigma )^2)}.
 \label{deltaEn}
\end{equation}
It is important to note that when the delta function is substituted with a Gaussian the detailed balance condition is only valid under approximation. This means that the OS definition of matrix $\mathbf{A}$ given in \cite{sparavigna:nc} and our definition, in Eq. \ref{ourA}, are not equivalent anymore. Our definition has the advantage to keep, for any finite $\sigma$ in Eq. \ref{deltaEn}, the symmetric and non-negative character  of the $\mathbf{A}$ matrix thanks to the symmetric definition of the scattering rate with the isotopic impurities given in Eq. \ref{isot1} and the replacement of $P^{\nu',\nu''}_{\nu}$ with $P_{\nu',\nu''}^{\nu}$. 

For diamond calculations: a smearing $\sigma=20$ cm$^{-1}$ along the ${\mathbf q'}$ mesh of $30 \times 30 \times 30$ has been found to lead to converged relaxation times (see Appendix \ref{gridsigma}).
For border scattering  we used a Casimir length $L=0.3$ cm and a shape factor $F=0.5$  \cite{sparavigna:prb65,born:book}.\\
A norm conserving pseudopotential\cite{troullier} with cutoff radii of 1.2 a.u. and core correction has been used for C.
The exchange correlation energy is calculated in the framework of the Local Density Approximation (LDA) \cite{ceperley}.
A plane-wave kinetic energy cutoff of 90 Ry and of 360 Ry for the charge density have been used. 
We used a  $8 \times 8 \times 8$ Monkhorst-Pack mesh in the BZ for the electronic k-point sampling. 

Anharmonic forces have been computed on a $4 \times 4\times 4$ q-point phonon grid on the BZ, Fourier interpolated with a finer $30 \times30 \times 30$ mesh for the Boltzmann calculations.

\section{Results and Discussion}
\begin{figure}[htp]
\includegraphics[width= 0.45\textwidth, angle=0]{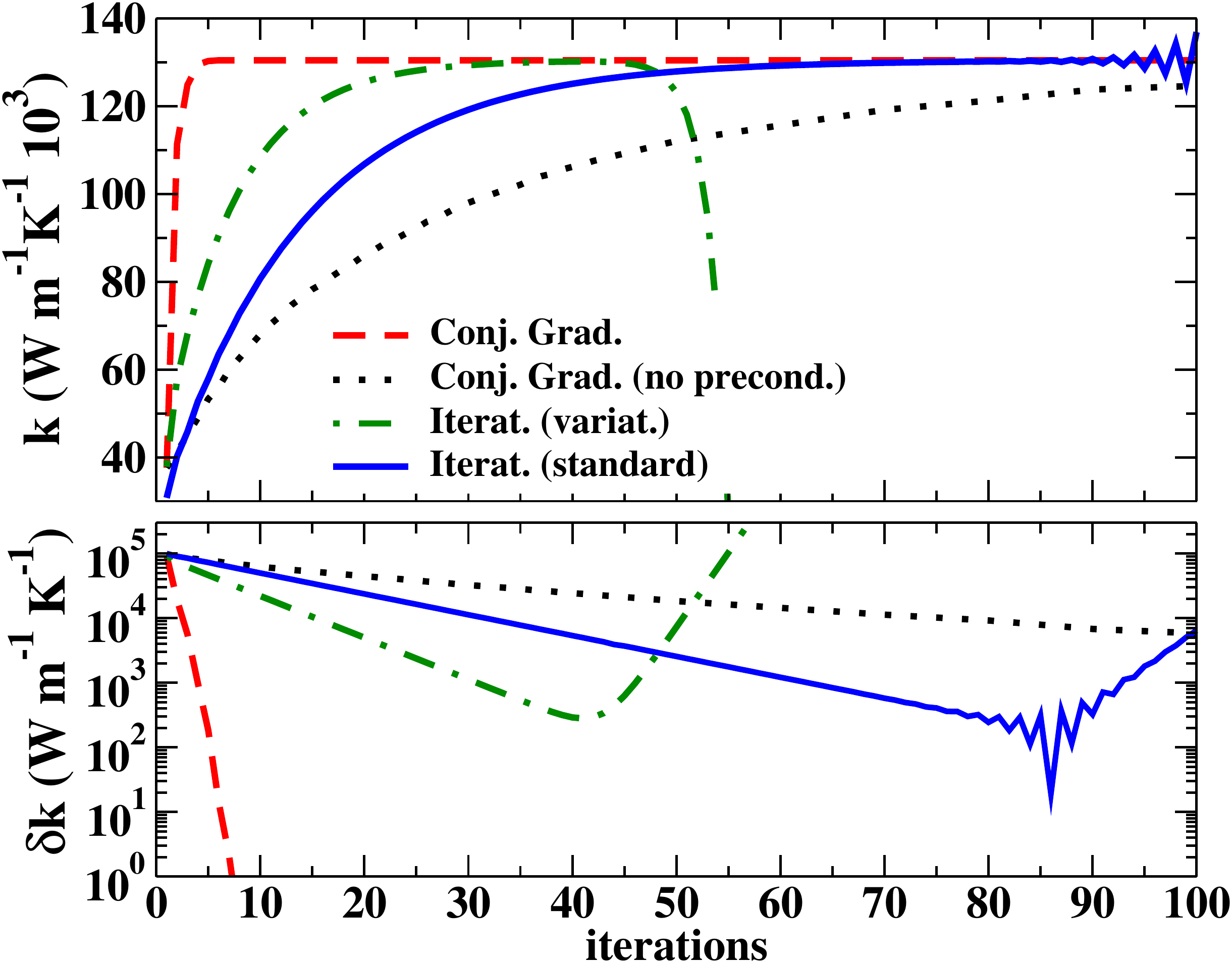}
\caption{(Color online). Lattice thermal conductivity of diamond at 100 K  (top panel)  and absolute error $\delta k$ (bottom panel) compared to the exact solution $k$ for: the iterative scheme in the Omini Sparavigna standard definition (solid line), the iterative scheme in the variational definition given in Eq. \ref{quadratic} (dash-dotted line),  and  the conjugate gradient method with (dashed line) and  without (dotted line) preconditioning, as a function of the order of iteration. }
\label{comparisonIT-CG}
\end{figure}

In Fig.\ref{comparisonIT-CG} a comparison between the convergence trend obtained via the OS iteration scheme or the conjugate gradient is 
reported for the case of bulk diamond at 100 K.
The OS standard iterative scheme shows a numerical instability after 77 iterations meaning that $\left(\mathbf{A}^{\mathrm{out}^ {-1}}  \mathbf{A}^{\mathrm{in}}\right) $ of Eq. \ref{os2}  has eigenvalues larger  than one in modulus. 
 This instability prevents the scheme from approaching the exact solution $k$ with a precision higher than $\sim 300$ W  m$^{-1}$ K$^{-1}$. A higher precision is achievable with the Conj. Grad. after just 4 iterations.

As expected, if the variational definition of $k$  (Eq. \ref{quadratic}) is used in the OS iterative scheme, half the number of iterations   are necessary to reach the same precision and the numerical instability appears after 41 iterations.
The convergence trend of the Conj. Grad. scheme without preconditioning is reported in the same graph to show how preconditioning is necessary to ensure a fast convergence. \\
We also considered an infinite diamond sample. The removal of the border effects does not change the Conj. Grad. convergence while the OS standard iterative procedure shows a numerical instability after 91 iterations with an error with respect to the exact solution of  $\sim 3000$ W  m$^{-1}$ K$^{-1}$.  This indicates how the OS  method becomes more unstable when scattering processes that do not conserve the crystal momentum (\emph{resistive} processes) are small.
Note that even with the most efficient Conj. Grad. approach, the CPU time required for obtaining the results shown in this paper has been two orders of magnitude larger than CPU time used for the IFCs ab initio calculation. Therefore the gain in speed up, with respect to the OS method, results in real speed up in the thermal conductivity calculations.

We have chosen for the comparison a temperature of 100 K 
 close to the maximum value of thermal conductivity obtainable in finite size diamond samples\cite{wei,broido:prb80,broido:underp}.  In this range of temperatures, where the U processes are a few and 
 the border effects are not dominant, it is important to have a stable algorithm able to well characterize the few scattering processes that drive the lattice thermal conductivity in order to obtain the correct result.\\
\begin{figure}[htp]
\includegraphics[width= 0.49\textwidth, angle=0]{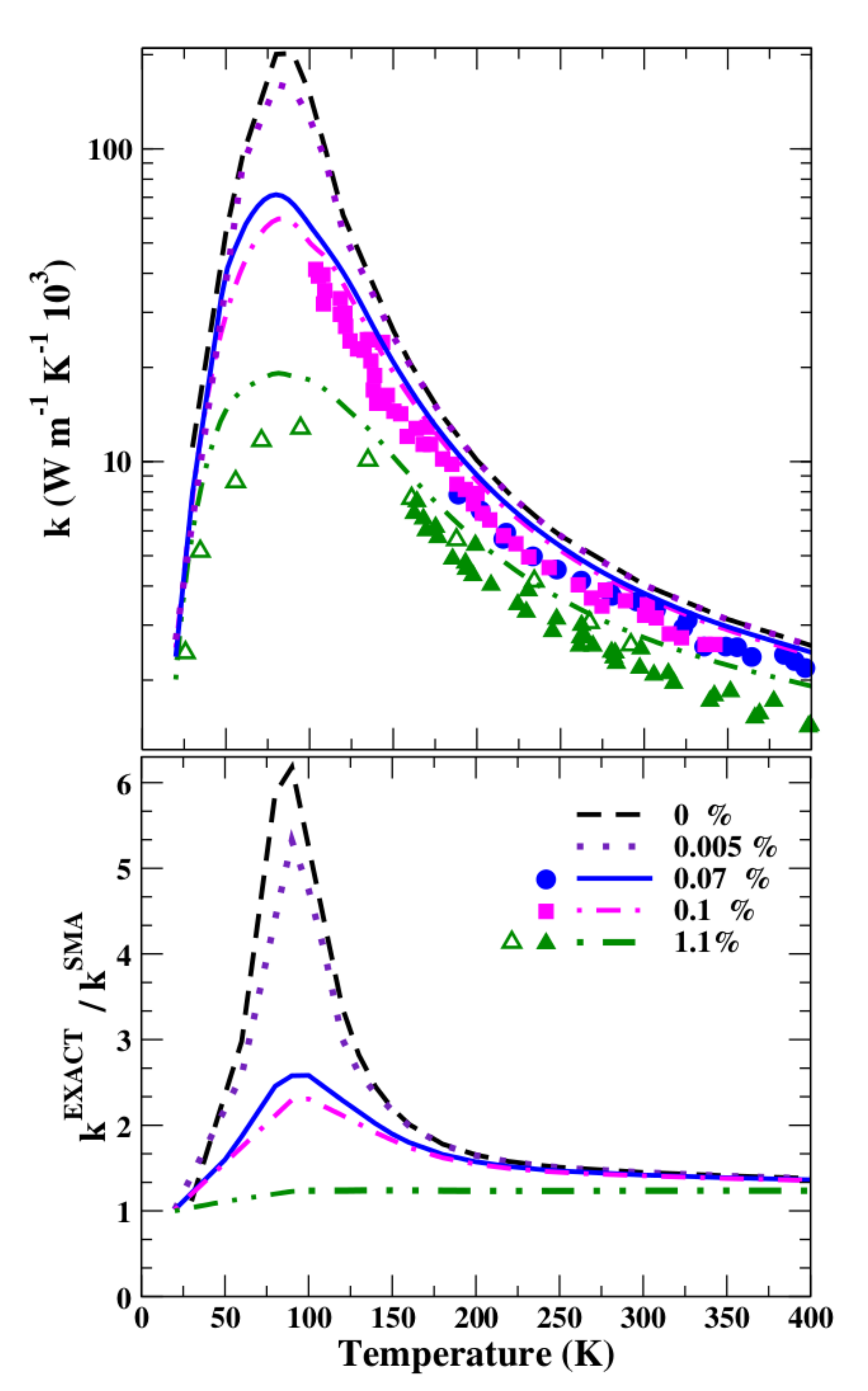}\\
\caption{(Color online). Lattice thermal conductivity of isotopically enriched and naturally occurring diamond as a function of temperature. (Top) Experimental values (circles \cite{wei}, squares \cite{olson:prb} and triangles \cite{wei,berman:jpc}) at different C$^{13}$ percentages
($0.07$\%, $0.1$\% and $1.1$\%) are compared with the results of our ab initio calculations (solid, dash-dotted and dash with two dots lines). As indication of the theoretical limit a dashed line for the case  in total absence of C$^{13}$ is reported.    (Bottom) Ratio between the exact and the SMA solution as a function of temperature at different C$^{13}$ percentages. }
\label{picture2}
\end{figure} \\
The top panel of Fig. \ref{picture2} compares the lattice thermal conductivity of isotopically enriched and naturally occurring diamond, obtained by solving exactly the
BTE equation, with the experimental results as a function of temperatures. The circles \cite{wei} and squares \cite{olson:prb}represent the measured values for isotopically enriched diamond with $99.93$\% C$^{12}$, $0.07$\%C$^{13}$ and $99.9$\% C$^{12}$, $0.1$\% C$^{13}$ respectively, while opened \cite{berman:jpc} and closed triangles\cite{wei} represent naturally occurring diamond with $98.9$\% C$^{12}$ and $1.1$\% C$^{13}$.
Our curves are in good agreement with experiments and with the previous theoretical results\cite{broido:prb80}, presented for T$\ge$ 150 K. As reported in Fig. \ref{picture2} there can be some discrepancies between different experiments due to the real dimension of the sample and to the presence of point defects, with the first playing a role in the low temperature regime, and the second becoming more relevant for higher temperatures. From Fig \ref{picture2} it is possible to infer that, in the case of naturally occurring diamond, the opened triangles\cite{berman:jpc} could be associated to samples with higher crystalline purity than the closed triangles\cite{wei} and, as expected, theoretical results, not considering the presence of structural defects, will always  be more in agreement with high purity samples.
In the same picture is also indicated with a dashed line the thermal conductivity in total absence of C$^{13}$.  This value gives a theoretical limit of the maximum lattice thermal conductivity reachable for an isotopically pure $\mathrm{C}^{12}$ sample. In the picture it is easy to notice that where the lattice thermal conductivity takes its maximum values $k_{0\%_{\mathrm{C}^{13}}} \simeq 3 k_{0.07\%_{ \mathrm{C}^{13}}}$. This means that there is still a significant increment in thermal conductivity achievable by growing more enriched diamond samples.

As the temperature increases, the values for the naturally occurring and isotopically pure samples become smaller.  This is due to the U scattering becoming stronger and consequently driving the thermal conductivity as the temperature increases. For temperatures lower than 80 K the border effects become dominant. \\  

The bottom panel of Fig.\ref{picture2} shows the ratio between the thermal conductivity obtained  by solving exactly BTE equation and by using the SMA as a function of temperatures. The lower the temperature and the less the C$^{13}$ abundance  the bigger becomes the ratio between the exact solution and the SMA solution. In other words,  the less are the events of scattering that do not conserve the momentum (i.e. Umklapp, isotopes and border scattering)  the less the SMA is able to give a good description of the process.
In Fig.\ref{picture2} is shown also the case with $99.995$\% C$^{12}$ and $0.005$\% C$^{13}$ as a further indication of how even small changes in the sample enrichment can give rise to sensible differences in the thermal transport properties of the material.

\begin{figure}[htp]
\includegraphics[width= 0.45\textwidth, angle=0]{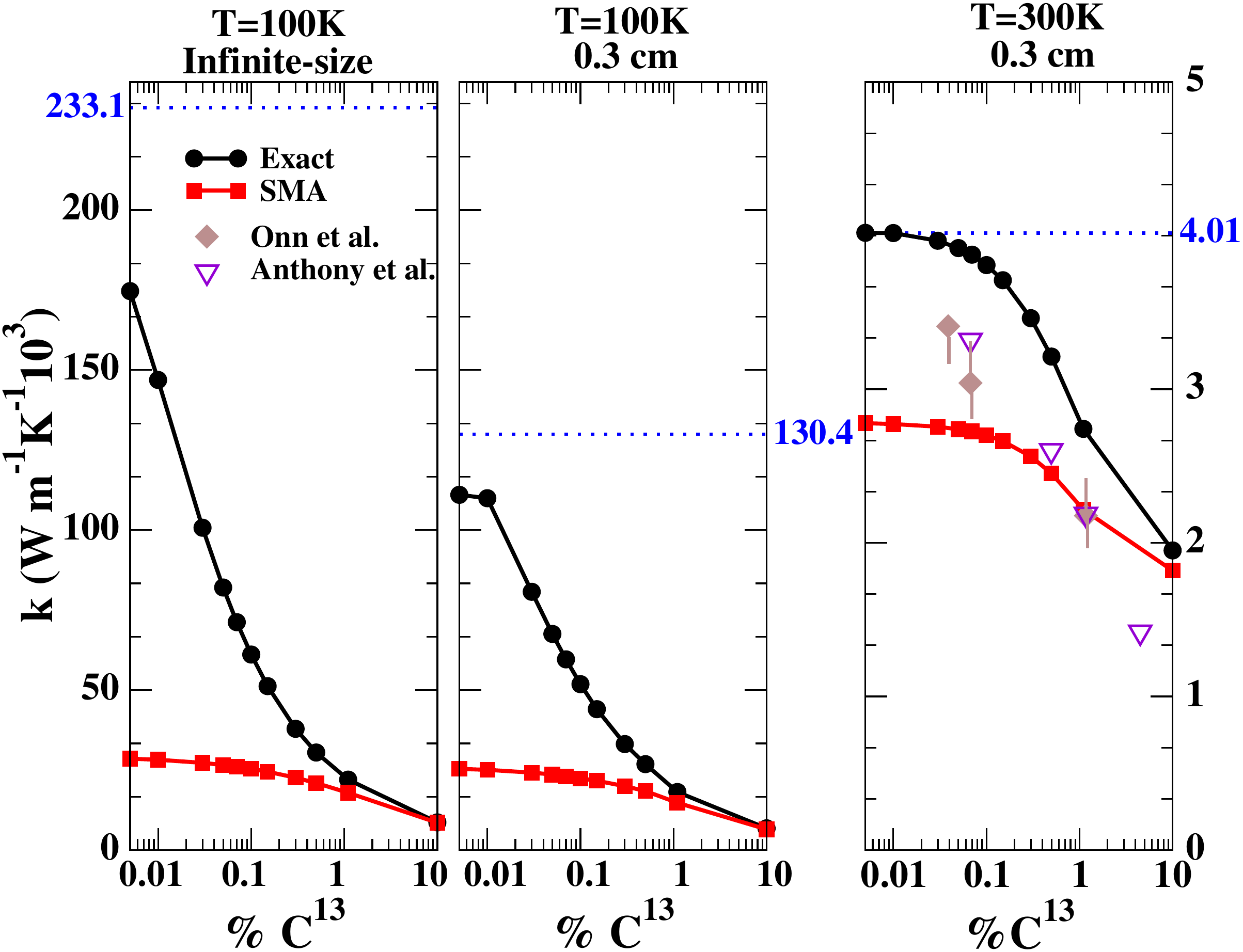}
\caption{(Color online) Diamond lattice thermal conductivity as a function of isotopic presence for an infinite-size diamond sample at 100 K (left) and for a sample with $L=0.3$ cm at 100 K (central) and 300 K (right). The values obtained by solving the BTE are indicated by circles while the SMA solutions by squares. The value of the thermal conductivity with $0$\% C$^{13}$ is represented, for each case, by a dotted horizontal line. At 300 K experimental values are indicated by diamonds\cite{onn}  and triangles \cite{anthony1,anthony2,anthony3}. } 
\label{picture3}
\end{figure}

In Fig.\ref{picture3} this last concept is more heightened. Diamond thermal conductivity is represented as a function of isotopic presence for two different temperatures $100$ K and $300$ K. 
At $T= 100$ K, for a finite-size diamond sample, the range of thermal conductivity explored by changing the percentage of C$^{13}$ from 0 to 1\% spans one order of magnitude while at 300 K the ratio is simply $1.5$. 
If, at 100 K, an infinite-size sample is considered, the thermal-conductivity dependence with respect to the isotopic content is enhanced. In particular, in the case of the finite-size sample, for isotopic percentages below  $0.01$ the lattice thermal conductivity  tends to a plateau, while in the infinite-size sample the curve does not show any deflection. 
This behavior  can be understood considering that Umklapp, isotope and border scattering are \emph{resistive} processes that make finite the value of thermal conductivity. At 100 K, where a few Umklapp scattering are activated and the border effects are non-dominant, the thermal conductivity value becomes very sensitive to even tiny variations of the isotopic content. This behavior is enhanced when the border effects are completely removed. At 300 K, as described above, the U processes are dominant with respect to the other non-momentum  conserving processes, so the lattice thermal conductivity shows a weaker dependency on the isotopic content. 
Equivalent experimental studies\cite{onn,anthony1,anthony2,anthony3} have been done at 300 K. The experimental points, as shown in Fig.4, present the same trend of our results but their values are slightly below our curve. Their lower value, as mentioned by the authors  themselves\cite{onn}, could arise from the level of crystallinity of the samples. In this respect, our calculations have the power to predict the effect of the isotopic content in the limit of perfectly crystalline samples. \\
\begin{figure}[htp]
\includegraphics[width= 0.45\textwidth, angle=0]{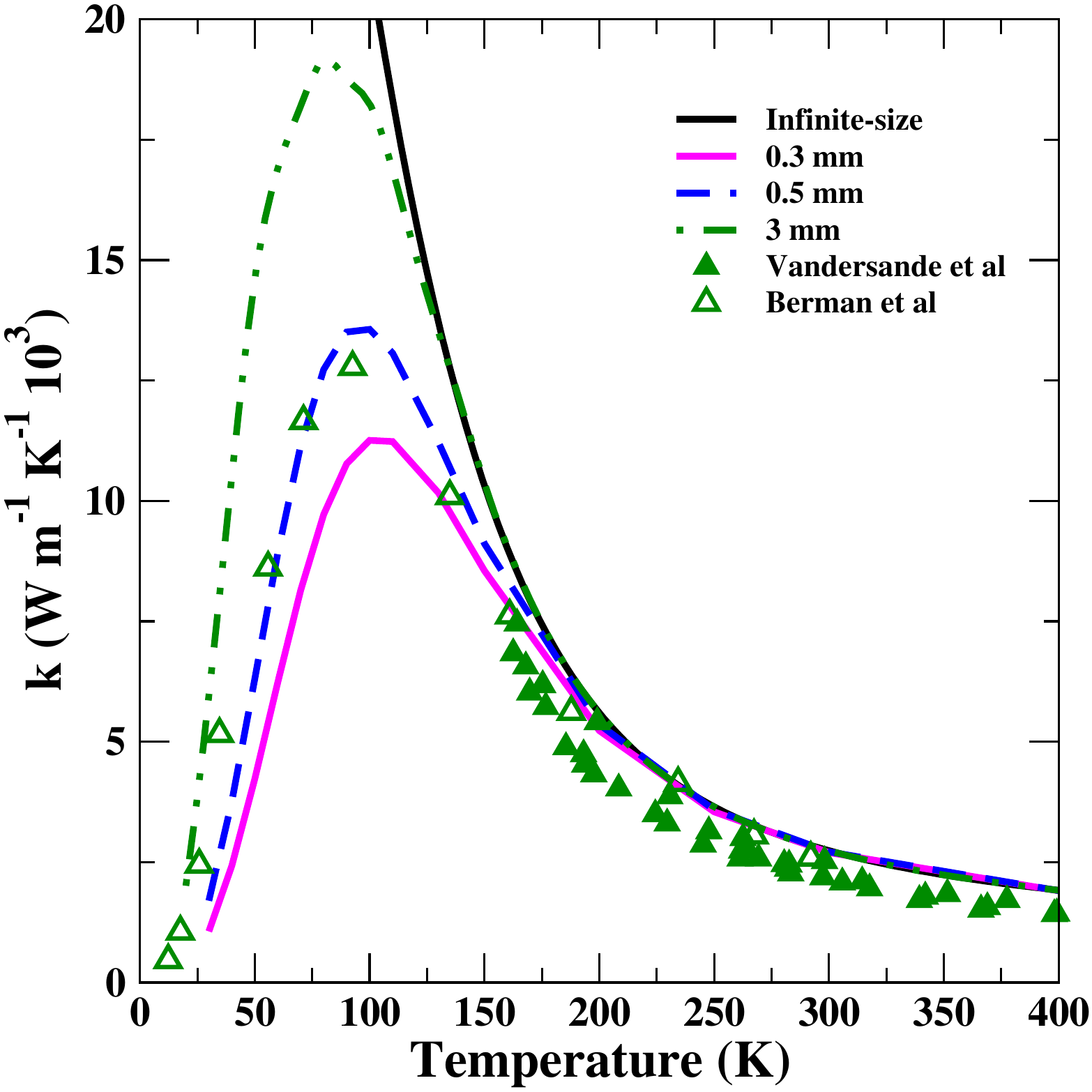}
\caption{(Color online) Diamond lattice thermal conductivity as a function of temperature for different sample dimensions in naturally occurring diamond. Experimental values are indicated by opened \cite{berman:jpc} and closed triangles\cite{wei}.}
\label{pictureSize}
\end{figure}
Furthermore, in order to show the role played by the dimension of the sample, in Fig.\ref{pictureSize} we report the lattice thermal conductivity, for naturally occurring diamond, as a function of temperature for different diamond-sizes. As described above the boundary scattering processes play a role in the low temperature regime. So, as it possible to see in Fig..\ref{pictureSize}, the larger is the diamond domain the higher is the maximum thermal conductivity achievable, with the limit of infinite $k$ for T$\to 0$ in infinite-size diamond. The theoretical curve obtained with $L=0.5$mm perfectly matches Berman et al.\cite{berman:jpc} results obtained on mm-size samples.

\section{Conclusions}

In this paper we have presented a new numerical approach for solving exactly the linearized BTE. We have shown how joining the variational principle approach and the resolution on a grid it is possible to always converge to the exact solution even in systems with very high thermal conductivity where resistive processes are weak.  Moreover the preconditioned conjugate gradient scheme with the line minimization assures a significantly faster convergence than the method previously proposed by Omini and  Sparavigna \cite{sparavigna:physb} with an equivalent computational cost per iteration, allowing to deal with larger grids than those accessible by the OS method. 

As a first application of our method we have computed the lattice thermal conductivity of isotopically enriched and naturally occurring diamond  by evaluating the harmonic and anharmonic IFCs fully ab initio in the framework of DFPT using a recent general implementation of the ``2n+1" theorem in the Quantum ESPRESSO package  combined with an exact solution of the linearized phonon BTE.   

In agreement with what previously shown in literature\cite{broido:prb80,sparavigna:prb65} we have demonstrated the severe inadequacy of the commonly used SMA in the range of temperature T$<$ 300 K for isotopically enriched diamond samples. 
In this range of temperatures, the lattice thermal conductivity shows a high sensitivity to the isotopic enrichment\cite{onn, olson:prb,berman:prb45}  and our calculations suggest that by growing more enriched diamond samples it is possible to achieve values of thermal conductivity up to three times larger than the commonly observed in isotopically enriched diamond samples with $99.93$\% C$^{12}$ and $0.07$ C$^{13}$.

\section{Acknowledgments}
This work was financed by ANR project \textsc{accattone}. Calculations were done at IDRIS (France), Project No.
096128 and CINES (France) Project \emph{imp6128}.

\appendix
\section{Properties of matrix A}
It is easy to prove that the matrix $\mathbf{A}$  is symmetric $A_{\nu,\nu'} - A_{\nu',\nu} = 0$
by using the properties:
 $P^{\nu''}_{\nu,\nu'} = P^{\nu''}_{\nu',\nu}$ and $P^{\mathrm{isot}}_{\nu,\nu'}=P^{\mathrm{isot}}_{\nu',\nu}$ in the definition of $A_{\nu,\nu'}$  given in Eq.\ref{ourA}.
It is also possible to prove that it is positive semi-definite. 
In order to show that, the matrix $\mathbf{A}$ in
Eq.\ref{ourA} can be written as:
\begin{equation}
\mathbf{ A}= \sum_{\nu,\nu'} P_{\nu,\nu'}^{\nu''} \mathbf{D}_{\nu,\nu'}^{\nu''} + \sum_{\nu,\nu'} P^{isot}_{\nu,\nu'} \mathbf{D}_{\nu,\nu'} + \sum_{\nu} P^{be}_{\nu} \mathbf{D}_{\nu} 
\end{equation}
where $\mathbf{D}_{\nu,\nu'}^{\nu''}$  is a matrix with all the element equal to zero apart those involving the triplets $\{ \nu,\nu',\nu'' \}$ 
\begin{center}
\begin{tabular}{l c r}
 $\:\: \:\: \:\: \:\:\: \:\: \:\: \:\:\: \:\: \:\: \:\:\: \:\: \:\: \:\: \nu$ & $\:\: \nu'$& $\: \nu''$\\
\end{tabular}
\vspace{-0.22cm}
\begin{equation}
\mathbf{D}_{\nu,\nu'}^{\nu''} =
\begin{tabular}{r}
$\nu\:$\\
$\:\nu'$ \\
$\:\:\nu''$ \\
\end{tabular}
\left(  
\begin{tabular}{l c r}
   \:1 &  \:1 & -1\\
   \:1 &  \:1 & -1 \\
 -1 & -1 &  1\\
\end{tabular}
\right )
\end{equation}
\end{center}
whose eigenvalues are: $0$,  $0$ and $3$. \\
For the part representing the elastic scattering with the isotopes:
 $\mathbf{D}_{\nu,\nu'}$  is a matrix with all the elements equal to zero apart those involving the couples $\{ \nu,\nu' \}$ 
\begin{center}
\begin{tabular}{l  r}
 $\:\: \:\: \:\: \:\:\:\: \:\:\: \:\: \:\: \:\:\: \:\: \:\:\: \nu$ & $\:\: \nu'$ \\
\end{tabular}
\vspace{-0.22cm}
\begin{equation}
\mathbf{D}_{\nu,\nu'} =
\begin{tabular}{r}
$\nu\:$\\
$\:\nu'$ \\
\end{tabular}
\left(  
\begin{tabular}{l  r}
   1 &  -1\\
 -1 &  1\\
\end{tabular}
\right )
\end{equation}
\end{center}
with eigenvalues $ 0$ and $2$.\\
Finally for the border effect, $\mathbf{D}_{\nu}$ is a matrix with all the elements equal to zero apart those involving $\{\nu,\nu\}$
\begin{center}
\begin{tabular}{l  }
 $\:\: \:\: \:\: \:\:\:\: \:\:\: \:\: \:\: \:\:\: \:\: \:\: \nu$ \\
\end{tabular}
\vspace{-0.26cm}
\begin{equation}
\mathbf{D}_{\nu} =
\begin{tabular}{r}
$\nu\:$
\end{tabular}
\left(  
\begin{tabular}{l  r}
   1 
\end{tabular}
\right )
\end{equation}
\end{center}

Since $P_{\nu,\nu'}^{\nu''}$, $ P^{\mathrm{isot}}_{\nu,\nu'}$  and  $P^{\mathrm{be}}_{\nu}$, are non-negative then the total matrix is positive semi-definite
 because  sum of positive semi-definite matrices.

\section{Phonon relaxation times}
When different events of scattering are present such as anharmonic scattering, scattering with isotopic impurities 
 and border effects the total  phonon relaxation time $\tau^{T}_{\bq j}$ is expressed by the Matthiessen's rule as:
\begin{equation}
(\tau ^{\mathrm{T}}_{\bq j})^{-1} =(\tau_{\bq j})^{-1} + (\tau^{\mathrm{be}}_{\bq j})^{-1}+ (\tau_{\bq j}^{\mathrm{isot}})^{-1} 
\end{equation}
where: 
 \begin{multline}
(\tau_{\bq j})^{-1}= 2 \Gamma_{\bq j}=\frac{\pi}{\hbar^2 N_0}\sum_{\bqp j',j''}|V^{(3)}(\bq j,\bqp j',\bqpp j'')|^2\\
	  \times \Big[2(\bar{n}_{\bqp j'}-\bar{n}_{\bqpp j''})\delta(\hbar \w_{\bq s}+\hbar \w_{\bqp j'}-\hbar \w_{\bqpp j''})+\\
	  (1+\bar{n}_{\bqp j'}+\bar{n}_{\bqpp j''})\delta(\hbar \w_{\bq j}-\hbar \w_{\bqp j'}-\hbar \w_{\bqpp j''})\Big]
\end{multline}
is the relaxation time due to the anharmonic scattering processes with $\Gamma_{\bq j}$ half width at half maximum of the corresponding phonon broadening;
while
\begin{equation}
(\tau^{\mathrm{be}}_{\bq j})^{-1} =\frac{c_{\bq j}} {L F} 
\end{equation}
is the relaxation time due to the border effects and
\begin{equation}
 (\tau_{\bq j}^{\mathrm{isot}})^{-1}{=} \frac{\pi}{2 N_0} \omega_{\bq j}^2 \sum_{\bqp j'} \delta (\hbar \omega_{\bq j}-\hbar \omega_{\bqp j'}) \\
   \sum_s  g^s_2 \left|\sum_{\alpha} z^{s \alpha^*}_{\bq j} z^{s \alpha}_{\bqp j'}  \right |^2 
\end{equation}
the relaxation time associated to the elastic scattering with isotopic impurities. \\

\section{Conjugate gradient method}

The conjugate gradient minimization \cite{num-rec} of Eq.\ref{fform} or Eq. \ref{tildefform}  requires the evaluation of the gradient $\mathbf{g}_i= \mathbf{A} \mathbf{f}_i - \mathbf{b}$
and a line minimization. Since the form is quadratic the line minimization can be done analytically and exactly.
Moreover the information required by the line minimization at  iteration $i$ can be recycled to compute the gradient at the next iteration $i+1$.
Starting with an the initial vector $\mathbf{f}_0= \mathbf{f}^{\mathrm{SMA}}$, initial gradient $\mathbf{g}_0=\mathbf{A}\mathbf{f}_0 -\mathbf{f}^{\mathrm{SMA}}$ and letting $\mathbf{h}_0= -\mathbf{g}_0$, the conjugate gradient method can be summarized with the
recurrence:
\begin{eqnarray}
\mathbf{t}_i&=&\mathbf{A} \mathbf{h}_i \\
  {\mathbf f}_{i+1} &=& {\mathbf f}_{i} - \frac{\mathbf {g}_{i} \cdot {\mathbf{h}_{i}} } {\mathbf{h}_{i} \cdot \mathbf{t}_i } \mathbf{h}_{i} \\
\mathbf{g}_{i+1}&=& \mathbf{g}_{i}-\frac{\mathbf {g}_{i} \cdot {\mathbf{h}_{i}} } {\mathbf{h}_{i} \cdot \mathbf{t}_i }\mathbf{t}_i\\
 \mathbf{h}_{i+1} &=& -\mathbf{g}_{i+1} + \frac{\mathbf{g}_{i+1} \cdot \mathbf{g}_{i+1}}{{\mathbf{g}_{i}} \cdot {\mathbf{g}_{i}} }  {\mathbf h}_{i} 
\end{eqnarray}
where $\mathbf{h}_i$ is the search direction and $\mathbf{t}_i$ is an auxiliary vector. Notice that each iteration requires only one application of the matrix $\mathbf{A}$ on the vector $\mathbf{h}_i$ as in the OS method. This is the computationally more demanding part of the conjugate gradient step.

\section{Grid and Smearing dependence} \label{gridsigma}
\begin{figure}[htp]
\includegraphics[width= 0.45\textwidth, angle=0]{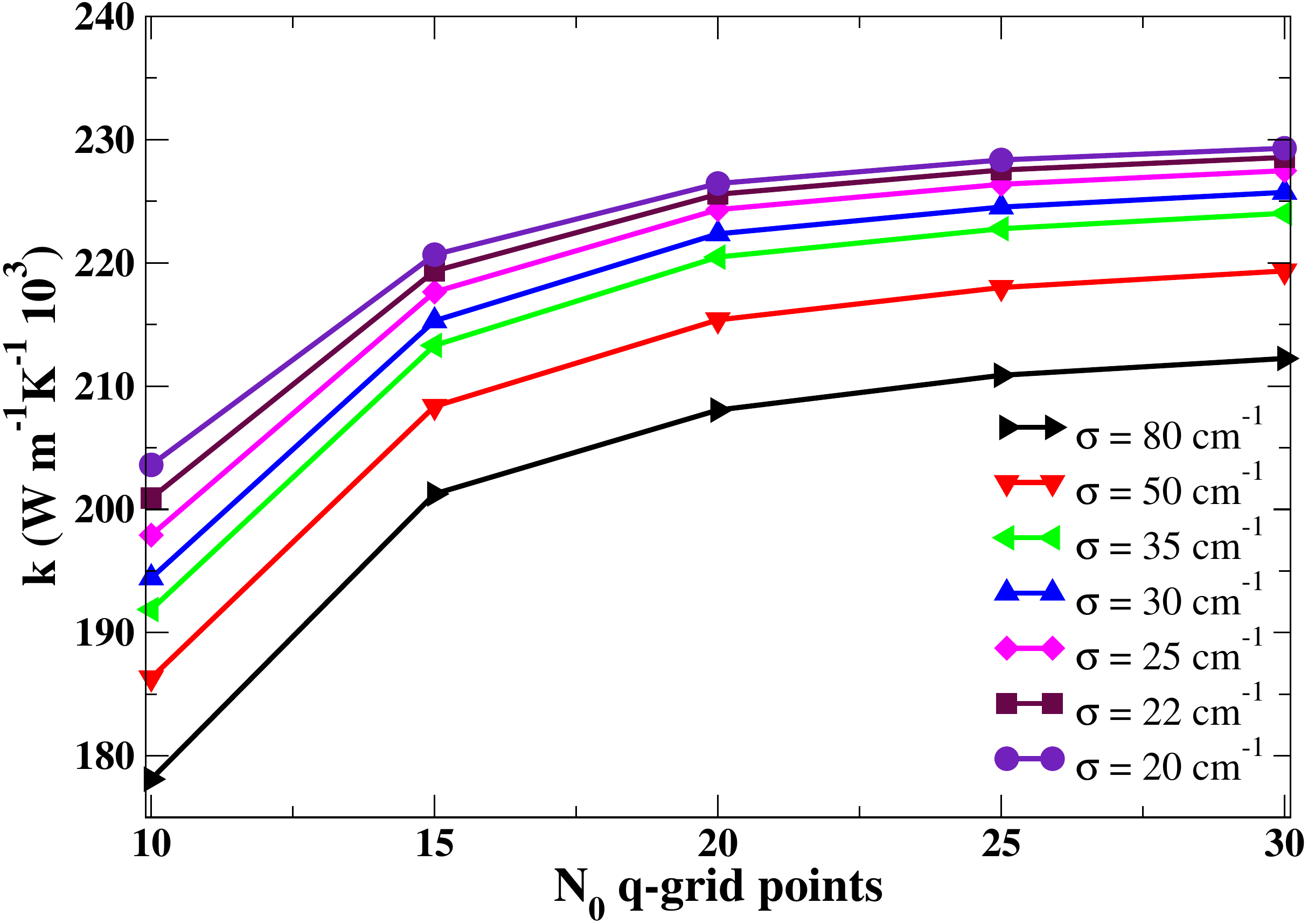}
\caption{(Color online)Diamond lattice thermal conductivity as a function of the number $N_0$ of the ${\mathbf q-}$grid points $(N_0\times N_0 \times N_0)$ at 100 K for different values (different curves)  of the energy-Gaussian smearing $\sigma$.}
\label{pictureCONV}
\end{figure}

In Fig.\ref{pictureCONV} Dependence of the lattice thermal conductivity at 100 K of an infinite-size diamond sample with $0$\% C$^{13}$ content, with respect to different energy-Gaussian smearing $\sigma$ and number $N_0$ of $\mathbf{q}$-grid points used. Notice that for the smaller grids the OS method shows numerical instability \cite{sparavigna:physb} from the very first steps while the Conj. Grad. does not present any slow down in convergence.

\newpage
\bibliography{biblio}
\end{document}